\documentclass[journal,transmag]{IEEEtran}
\IEEEoverridecommandlockouts
\usepackage{cite}
\usepackage{amsmath,amssymb,amsfonts}
\usepackage{graphicx}
\usepackage{textcomp}
\usepackage{xcolor}
\usepackage{dsfont}
\usepackage{caption}
\usepackage{subcaption}
\usepackage{booktabs,makecell}
\usepackage{booktabs}
\usepackage{makeidx}
\usepackage{algorithm}
\usepackage[noend]{algpseudocode}
\usepackage{amsthm}
\usepackage{tikz}
\usepackage{adjustbox}
\usepackage{color}
\usepackage{algcompatible}
\usepackage{arydshln}
\usepackage{bbm}
\usepackage{multirow}

\usepackage{bbm}

\def\BibTeX{{\rm B\kern-.05em{\sc i\kern-.025em b}\kern-.08em
		T\kern-.1667em\lower.7ex\hbox{E}\kern-.125emX}}
\begin{document}
	
	\title{Reduced Complexity Neural Network Equalizers for Two-dimensional Magnetic Recording\thanks{*Core Technical Contributor.}
	}
	
	\author{
	\IEEEauthorblockN{Ahmed Aboutaleb and  Nitin Nangare*}
	\IEEEauthorblockA{Marvell Technology Inc., Santa Clara, CA 95054, USA, aaboutaleb@marvell.com, nitinn@marvell.com}%
}
	
	
	\IEEEtitleabstractindextext{%
	\begin{abstract}
This paper investigates reduced complexity neural network (NN) based architectures for equalization over the two-dimension magnetic recording (TDMR) digital communication channel for data storage.
We use realistic waveforms measured from a hard disk drive (HDD) with TDMR technology. We show that the multilayer perceptron (MLP) non-linear equalizer achieves a $10.91\%$ reduction in bit error rate (BER) over the linear equalizer with cross-entropy-based optimization. However, the MLP equalizer's complexity is $6.6$ times the linear equalizer's complexity. Thus, we propose reduced complexity MLP (RC-MLP) equalizers. Each RC-MLP variant consists of finite-impulse response filters, a non-linear activation, and a hidden delay line. A proposed RC-MLP variant entails only $1.59$ times the linear equalizer's complexity while achieving a $8.23\%$ reduction in BER over the linear equalizer.
		
	\end{abstract}
	
	}
		\maketitle
		
		\section{Introduction}
\begin{figure}[t]
			\centering
			\includegraphics[width=0.35\textwidth]{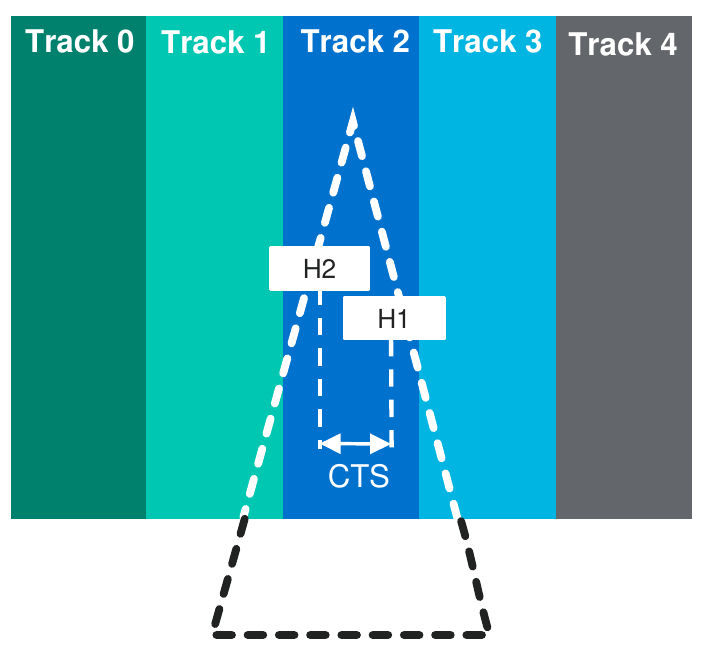}
			\caption{{\small TDMR reader. The reader has two read heads to enable the compensation of inter-track interference (ITI) due to decreased track pitch in TDMR. The cross-track separation (CTS) represents the distance between the two read heads H1 and H2.}}
			\label{TDMR_Diagram}
		\end{figure}

Hard disk drives (HDDs) will continue to be cost-competitive as user bit density increases. Data is stored on HDDs through magnetic recording digital communication channels. The conventional one-dimensional magnetic recording (1DMR) channel stores bits along a single track with sufficient spacing between tracks to prevent inter-track interference (ITI). Two-dimensional magnetic recording (TDMR) increases the density of information by decreasing the track pitch, resulting in significant ITI in the readback waveform \cite{The_Feasibility_MR_at_10_Tb_Wood}. TDMR currently uses two or more closely spaced read heads positioned in the HDD slider to help compensate for the increased ITI. Figure \ref{TDMR_Diagram} illustrates a TDMR reader with two heads.
 \subsection{Trellis-based Viterbi Detection for TDMR}
The read heads measure the waveforms that are read back from the magnetic media \cite{vasic2004codingandSignalProcessingforMagneticRecordingSystems}. During readback, such waveforms are passed through a low-pass anti-aliasing filter. To acquire discrete-time samples, an analog-to-digital converter (ADC) samples the filter's output at an appropriate rate. The inter-symbol interference (ISI) in the ADC samples spans many bits in the down-track direction. Most commonly, bits are estimated using a trellis-based detector. With the Viterbi algorithm (VA), maximum likelihood (ML) detection can be achieved with a trellis-based approach at an acceptable level of complexity \cite{TheViterbiAlgorithmForney1973}. Nevertheless, as the ISI length increases, the number of states in the trellis detector increases exponentially. In addition, the number of computations per bit estimate is directly proportional to the number of trellis states. Because of the ISI/ITI in TDMR, the number of trellis states may be high, which leads to a high level of implementation complexity in trellis-based detection.

Moreover, the canonical VA detector assumes Gaussian noise and linear ISI/ITI in the coded bits in order to achieve optimality in the ML sense. In practice, neither of these assumptions is generally true. In fact, the readback waveforms contain data-dependent noise, partial erasures, non-linear ISI/ITI, and jitter noise \cite{Data_Storage_Channel_Equalization_Neural_Networks_Nair,Nonlinear_Equalization_TDMR_Using_NNs_Shen}. 
To shorten the ISI/ITI and whiten the noise, an equalizer is used before the trellis detector. Typical data recovery systems employ a 2D-linear minimum mean square error (2D-LMMSE) equalizer followed by a VA detector \cite{Data_Storage_Channel_Equalization_Neural_Networks_Nair}. To simplify the implementation of the 2D-LMMSE equalizer, the equalizer is implemented as a finite impulse response (FIR) filter. Accordingly, the VA detector operates on the output samples of the equalizer while assuming a manageable number of trellis states. 

\subsection{Sub-optimality of the Linear Equalizer}

However, the 2D-LMMSE equalizer is optimal in the mean-square error (MSE) only if the noise is Gaussian and stationary and if the channel is linear in terms of the written bits \cite{Lessons_Estimation_Theory_Mendel}. Furthermore, an equalizer trained to minimize the MSE may not result in the best detector bit error rate (BER), which is the desired figure of merit. Further, as the storage channel areal density increases, non-linear impairments, residual ISI and ITI, and signal-dependent noise become more significant \cite{Data_Storage_Channel_Equalization_Neural_Networks_Nair}. Therefore, the actual channel conditions differ substantially from those required for optimality. In short, the 2D-LMMSE equalizer is not optimal in the desired metric (BER), and its MSE optimality conditions are not met in practice.

	\subsection{Relevant Works on Neural Networks for Magnetic Recording}
NNs are capable of compensating for various types of impairments, so long as they are trained with appropriate representative data. Recent studies have demonstrated that neural networks (NNs) can provide improved performance in high-density magnetic recording channels, despite the presence of higher amounts of media noise than in conventional 1D channels. The use of neural networks significantly improves the overall system performance of the 1DMR \cite{Data_Storage_Channel_Equalization_Neural_Networks_Nair}, the TDMR \cite{Nonlinear_Equalization_TDMR_Using_NNs_Shen,DNN_APP_Detector_TDMR_Shen,DeepNeuralNetworkMediaNoisePredictorTurboDetectionSystemfor1Dand2DHighDensityMagneticRecording_Sayyafan, Study_NN_Equalization_TDMR_Luo}, and the multilayer magnetic recording (MLMR) \cite{DeepNeuralNetwork-basedDetectionandPartialResponseEqualizationforMultilayerMagneticRecording} systems.  

In \cite{Data_Storage_Channel_Equalization_Neural_Networks_Nair}, Nair and Moon propose using a multilayer perceptron (MLP) as an equalizer for high-density 1DMR channels. The MLP is a fully connected feed-forward neural network with one or more hidden layers (cf. \cite{DeepLearning_Goodfellow}). Their results demonstrate that, as a non-linear equalizer, the MLP outperforms the conventional linear equalizer in terms of MSE and BER. Sayyafan {\it et al.} have proposed to integrate a convolutional neural network (CNN) with Bahl–Cocke–Jelinek–Raviv (BCJR) in order to iteratively estimate and cancel the data-dependent media noise \cite{DeepNeuralNetworkMediaNoisePredictorTurboDetectionSystemfor1Dand2DHighDensityMagneticRecording_Sayyafan}. In both 1DMR and TDMR channels, their proposed system achieves significant information density gains over conventional pattern-dependent noise prediction (PDNP) systems. An investigation conducted by Luo {\it et al.} demonstrates that the NN equalizer outperforms the 2D-LMMSE equalizer in a TDMR channel \cite{Study_NN_Equalization_TDMR_Luo}. In addition, Shen {\it et al.} have presented a 2D-LMMSE equalizer followed by a CNN detector for application in TDMR \cite{DeepNeuralNetwork-basedDetectionandPartialResponseEqualizationforMultilayerMagneticRecording}. This system outperforms a conventional system comprised of a 2D-LMMSE equalizer along with a 2D-BCJR and a 2D-PDNP. 
In \cite{DeepNeuralNetwork-basedDetectionandPartialResponseEqualizationforMultilayerMagneticRecording}, Aboutaleb {\it et al.} have proposed CNN-based equalization and detection systems for MLMR channels suffering from severe data-dependent noise. 
Compared to conventional 2D-LMMSE equalizer-VA systems, CNN-based systems achieve significant density gains.
Shen and Nangare have proposed an NN equalizer followed by a VA in which the parameters of the NN adapt to minimize the cross-entropy (CE) \cite{Nonlinear_Equalization_TDMR_Using_NNs_Shen}. By adapting the equalizer's parameters to minimize the CE loss, the authors demonstrate lower detector BERs compared to MSE-adapted equalizer parameters. Their study suggests that CE-based optimization correlates better with improved BERs, compared with MSE-based optimization. Further, their study confirms that NN equalizers can be used to compensate for the various impairments associated with high-density magnetic recording systems.

\subsection{High Complexity of Neural Networks for Equalization and Detection}

Despite the improvements in performance reported in \cite{DNN_APP_Detector_TDMR_Shen,DeepNeuralNetworkMediaNoisePredictorTurboDetectionSystemfor1Dand2DHighDensityMagneticRecording_Sayyafan, Study_NN_Equalization_TDMR_Luo,DeepNeuralNetwork-basedDetectionandPartialResponseEqualizationforMultilayerMagneticRecording}, the mentioned NN-based equalizers require a much higher implementation complexity than the linear equalizer baseline. Indeed, the high complexity of NN-based methods hinders practical implementation. For example, among the lowest complexity NN equalizers proposed by previous studies, the complexity of the MLP in \cite{Nonlinear_Equalization_TDMR_Using_NNs_Shen} is about $6.6\times$ the complexity of the 2D-LMMSE. 
Moreover, recent studies on NN-based equalizers have not proposed any architectures with complexities that are $3\times$ or lower than the complexity of the baseline linear equalizer. 

\subsection{Overview of the Proposed Equalizers}
We propose four variants of a reduced complexity MLP (RC-MLP) architecture to facilitate practical implementation. RC-MLP contains FIR filters, hidden delay lines, and non-linear activation functions; these components can easily be implemented in practice. We demonstrate that RC-MLP architectures deliver an excellent balance between performance and complexity. Among the proposed variants, one architecture offers the best performance-complexity trade-off. 

The novel contributions of this paper are summarized as follows.

\begin{enumerate}
\item We investigate reduced complexity neural network architectures for equalization over TDMR channels. We propose four variants of RC-MLP that achieve most of the performance gains of high-complexity and high-performance MLPs.
\item Additionally, we consider candidate equalizer architectures based on radial basis function neural networks (RBFNNs). Our study examines the performance and complexity of these architectures. 
\item Our experiments use data and readback waveforms measured from an HDD to evaluate the performance of the proposed and baseline methods. The data is obtained from a hard disk drive with TDMR technology. Performance and complexity are compared for each method. Then, we identify the architecture with the best balance between performance and complexity.	
\end{enumerate}
\section{System Model}
	This paper uses actual HDD waveforms with TDMR technology using two read heads for training and testing. The raw unequalized waveforms are sampled synchronously at one sample per bit per read head. 
	In the next paragraph, we summarize a channel model that approximates the TDMR channel, where details can be found in \cite[Sec. II]{Nonlinear_Equalization_TDMR_Using_NNs_Shen} and \cite{MultidimensionalSignalProcessingandDetectionforStorageSystemswithDatadependentTransitionNoise}.
	
	\subsection{Summary of a Channel Model}
	Let $\mathbf{u}=u_n\in \{-1,+1\}$ be a binary input sequence to be written on the media. 
	Define the transition sequence $\mathbf{b}$ such that $b_n \triangleq (u_n - u_{n-1})/2$. 
	Let $h(t,w)$ denote the 2D channel response, modeled using the erf$(\cdot)$ function as in \cite{Nonlinear_Equalization_TDMR_Using_NNs_Shen, GeneralizedPRTargetswithPerpendicularRecording, MultidimensionalSignalProcessingandDetectionforStorageSystemswithDatadependentTransitionNoise}. Then, the continuous-time readback waveform $r_a(t)$ is given by:
	\begin{align}
		r_a(t) = \sum_{n} b_nh(t-nT+\Delta t_n, w+\Delta w_n) + n(t),
	\end{align}
	where $T$ is the symbol interval, $\Delta t_n$ is the down-track jitter noise, modeled as a Gaussian random process using a truncated Gaussian distribution such that $|\Delta t_n| < T/2$, $\Delta w_n$ is the cross-track jitter noise modeled as a truncated Gaussian distribution, and $n(t)$ is the AWGN, which models the reader electronics noise. 
	
	Let $p(t,w) \triangleq h(t,w) - h(t-T,w)$ denote the channel bit-response. Then, the equivalent discrete-time channel model is given by \cite{Nonlinear_Equalization_TDMR_Using_NNs_Shen},:
	\begin{align}
		r^{\langle l\rangle}_n = (\mathbf{p}^{\langle l\rangle}*\mathbf{u})_n + n_n^{\langle l\rangle},\quad l=0,1,
	\end{align}
	where $*$ denotes discrete-time convolution, $\mathbf{p}^{\langle l\rangle}$ is the 2D response for reader $l$. 
	It is worth noting that more accurate data sets can be synthesized using the grain flipping probability model to produce the magnetizated bit cells as $\mathbf{u}$ by simulating media-data interactions \cite{ChannelModelsandDetectorsforTDMR_Chan,AD_Capability_Dual_Structure_Media_MAMR_Greaves}. This approach has been demonstrated to generated readback waveforms that closely match with readback measurements from HDDs. The equalizer's parameters can be adapted using the CE or MSE loss.
	
	\begin{figure}[t]
		\centering
		\includegraphics[width=0.48\textwidth]{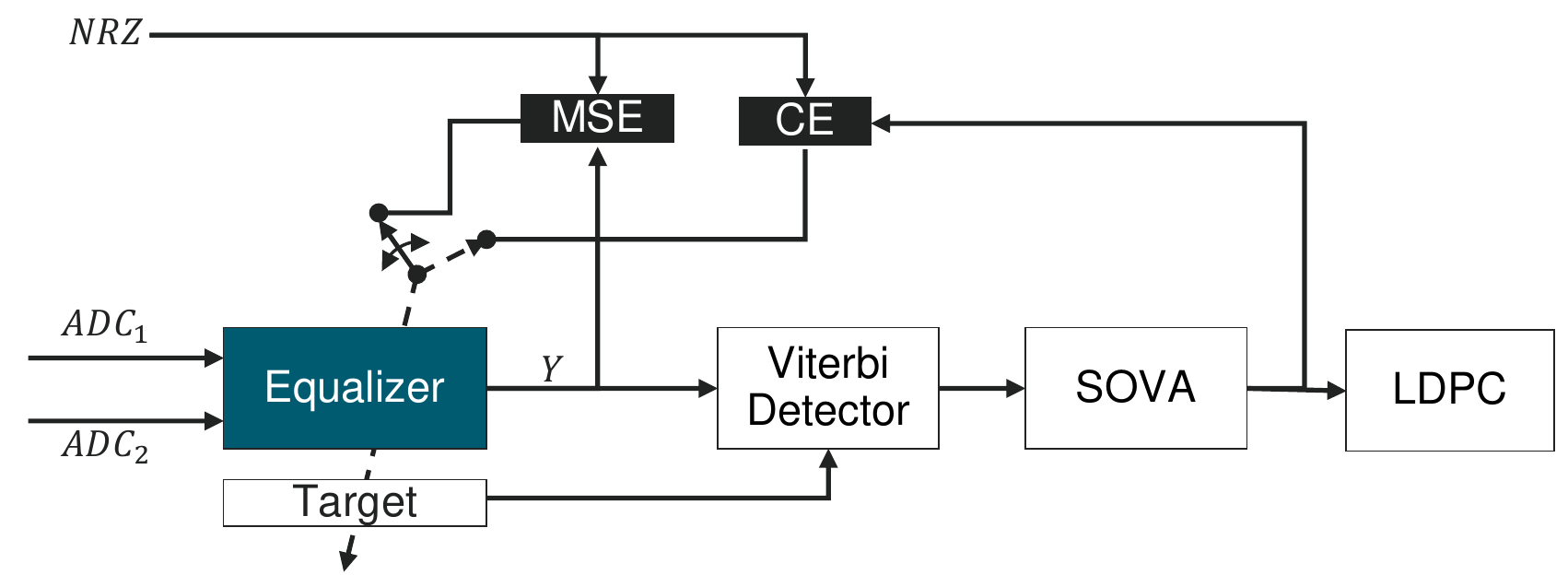} 
		\caption{{\small Equalizer-detector system. The equalizer accepts the readback ADC samples and outputs the PR signal $\mathbf{y}$. The Viterbi detector, along with the SOVA, compute the soft-bit estimate $\mathbf{p}_0$.}}
		\label{System_diagram}
	\end{figure}
\subsection{General Equalizer-Trellis Detector System}
	Fig.~\ref{System_diagram} shows the equalizer-detector system. The equalizer processes the ADC samples from the readers. We assume one reader per track and that the outputs of two readers are processed simultaneously. The equalizer's output is passed to the SOVA detector \cite{AViterbiAlgorithmwithSoftDecisionOutputsandItsApplication}, which computes soft bit estimates as LLRs. The LLRs are then fed to the channel decoder to recover the information bits. The equalizer's trainable parameters and target are adjusted using the MSE or the CE criteria.

	
	%

	\section{Adaptation}
	
	Let the equalizer's output be denoted by $y_n$ and the noise-free PR target signal by $\hat{y}_n$. The equalizer output $y_n$ depends on the equalizer's architecture design and its learnable parameters, which include the weights $\mathcal{W}$ and biases $\mathcal{B}$. The equalizer's design is discussed in Sections \ref{NNEqualizers} and \ref{ProposedNNEqualizers}. The noise-free PR signal is given by:
	\begin{align}
		\hat{y}_n = (\mathbf{g}*\mathbf{u})_n,
	\end{align}
	where $\mathbf{g}$ represents the PR target, and $\mathbf{u}$ represents the written bit sequence.
		\subsection{Mean-Squared Error Loss Function}
	 For a length-$N$ minibatch, the average MSE $J_{\text{MSE}}$ is commuted as:
	\begin{align}
		J_{\text{MSE}} = \dfrac{1}{N}\sum_{n=0}^{N-1}(\hat{y}_n - y_n )^2.
	\end{align}
	For MSE adaptation, the optimization problem is given by:
	\begin{subequations}
		\label{constrainedMSE_RC_MLP}
		\begin{alignat}{4}
			&\!\underset{\mathcal{W}, \mathcal{B}, \mathbf{g}}{\text{minimize}}   &\quad& J_{\text{MSE}} \\
			&\text{subject to} &      & \mathbf{a}^T\mathbf{g}=1, \label{monic_constraint1}
		\end{alignat}
	\end{subequations} 
	where \eqref{monic_constraint1} is the monic constraint (MC) on the target which sets the first tap of the target to one, i.e., $\mathbf{a}=[1,0,0,\ldots,0]$. In MSE adaptation, including the MC prevents the target coefficients from converging to the trivial solution of $\mathbf{g}=\mathbf{0}$ and has been shown to improve the BER performance \cite{Equalization_for_ML_Detectors_Moon}.
	
	\subsection{Cross-Entropy Loss Function}
	Consider estimating the $n$th bit $u_n$. Its estimate is denoted by $\hat{u}_n$. For CE adaptation, $\hat{u}_n$ is obtained by passing the equalizer's output to the VA and computing the soft decisions, in the form of log-likelihood ratios (LLRs), using the soft output VA (SOVA) \cite{AViterbiAlgorithmwithSoftDecisionOutputsandItsApplication}. Let $ \mathds{1}_{u_n=i}$ denote the indicator function such that $ \mathds{1}_{u_n=i} = 1$ if $u_n=i$, $i\in \{-1,1\}$ (and zero otherwise), and define $p_{0,n} \triangleq \Pr\{\hat{u}_n = -1\}$.  
	Then, for the $n$th bit, the CE loss is computed as:
	\begin{align}
		\mathcal{H}\{u_n, \hat{u}_n\} =
		- \mathds{1}_{u_n=-1}\log(p_{0,n}) -\mathds{1}_{u_n=1}\log( 1 - p_{0,n}). \label{CrossEntropy}
	\end{align}
The estimate LLR$_n$ for the $n$th bit provided by the SOVA is defined as\cite{AViterbiAlgorithmwithSoftDecisionOutputsandItsApplication,Nonlinear_Equalization_TDMR_Using_NNs_Shen}:
	\begin{align}
		\text{LLR}_n &\triangleq \log\Bigg(\dfrac{\Pr\{\hat{u}_n=1\} }{\Pr\{\hat{u}_n=-1\}}\Bigg) \\
		&= \log\Bigg( \dfrac{1-p_{0,n} }{p_{0,n}}\Bigg). 
	\end{align} 
	Then, $p_{0,n}$ is computed as:
	\begin{align}
		p_{0,n} = \dfrac{1}{1+e^{\text{LLR}_n}}. \label{ProbabilityEstimate}
	\end{align}
	By substituting \eqref{ProbabilityEstimate} in \eqref{CrossEntropy}, it can be shown that the CE is given by:
	\begin{align}
			\mathcal{H}\{u_n, \hat{u}_n\} =\log\Big( 1 + e^{-{u_n\text{LLR}_n}} \Big).
	\end{align}
	The objective function to minimize is the average CE loss computed over a minibatch of length $N$ as:
	\begin{align}
		J_{\text{CE}} = \dfrac{1}{N}\sum_{k=0}^{N-1} 	\mathcal{H}\{u_n, \hat{u}_n\}.
	\end{align}
	Hence, the optimization problem can be written as:
	\begin{alignat}{4}
		&\!\underset{\mathcal{W}, \mathcal{B}, \mathbf{g}}{\text{minimize}}   &\quad& J_{\text{CE}}. \label{constrainedCE}
	\end{alignat}
	The optimization problems in \eqref{constrainedMSE_RC_MLP} and \eqref{constrainedCE} are solved using the back-propagation algorithm with stochastic gradient descent (SGD) \cite[Ch. 8]{DeepLearning_Goodfellow}. 
	
	Adapting the equalizer's parameters using the CE criterion is equivalent to maximum likelihood adaptation since the BER and soft-bit information are jointly improved. Thus, for any given equalizer structure CE adaptation outperforms MSE adaptation in terms of BER \cite{Nonlinear_Equalization_TDMR_Using_NNs_Shen}.

	\section{Neural Network Equalizers}
	\label{NNEqualizers}
The MLP and RBFNN are universal approximators under mild conditions
 \cite{MLP_UniversalApproximators,UniversalApproximationRBFNN}. As such, they are capable of approximating any continuous input-output mapping provided they have sufficient hidden nodes and appropriate data to train them. The universality of the MLP and RBFNN makes them ideal candidates for mitigating the impairments and data-dependent noise observed in high density magnetic recording.   
	
	\subsection{Multilayer Perceptron}
	
	\begin{figure}[t]
		\centering
		\includegraphics[width=0.4\textwidth]{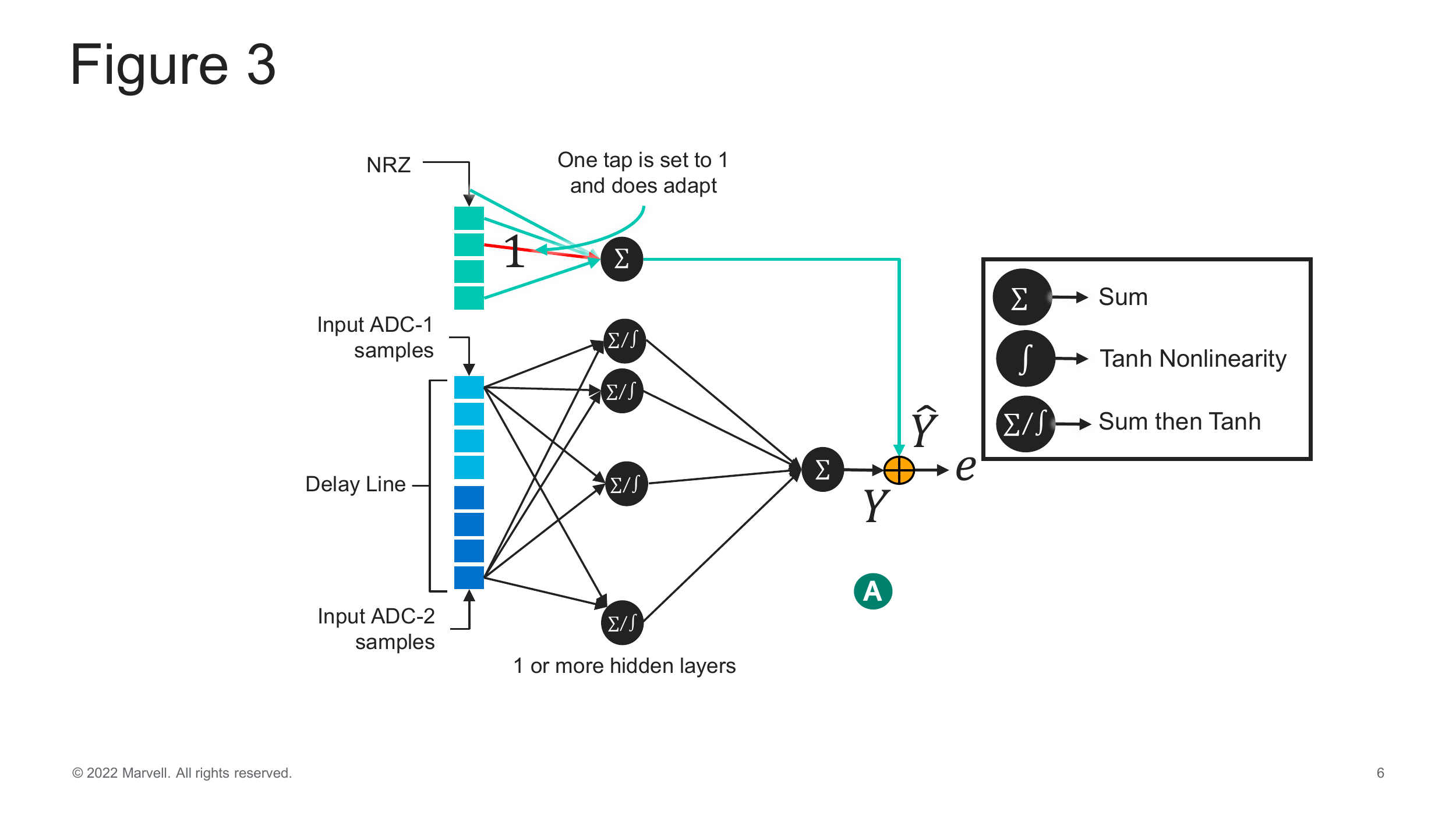}
		\caption{{\small Multilayer perceptron (MLP). The inputs are delayed ADC samples over a sliding window. The connections here represent a fully connected layer consisting of a matrix with dimensions consistent with the input length and the designed number of hidden outputs. The final equalizer output is an affine combination of all hidden outputs. The notation $\sum/\int$ represents a summation followed by a hyperbolic tangent activation. }}
		\label{MLP_diagram}
	\end{figure}
	
	Fig.~\ref{MLP_diagram} shows the architecture of the MLP equalizer. To estimate the $n$th bit in the downtrack direction, the MLP uses the ADC samples observed over a sliding window of size $2M+1$ for each reader centered around the $n$th readback sample. Let $\mathbf{r}_n \triangleq [\mathbf{r}_n^{\langle 0\rangle},\mathbf{r}_n^{\langle 1\rangle}]$ denote such ADC samples from the TDMR reader centered around the $n$th sample, where the ADC samples per reader are defined as: $\mathbf{r}_n^{\langle l\rangle} \triangleq [r^{\langle l\rangle}_{n-M}, r^{\langle l\rangle}_{n-M+1},\ldots, r^{\langle l\rangle}_n, r^{\langle l\rangle}_{n+1}, \ldots,r^{\langle l\rangle}_{n+M} ]^T $, $l=0,1$. The MLP contains a matrix of weights $\mathbf{W}=[w_{i,j}]\in \mathbb{R}^{(4M+2) \times K}$ that multiplies the vectorized input samples $\mathbf{r}_{n,\text{vec}} = \text{vec}(\mathbf{r}_n)$ to provide $K$ intermediate outputs. After adding the biases $b_{0,k}$, $k=1,\ldots,K$, to the intermediate outputs, the hidden outputs are obtained by applying an element-wise non-linear activation function $\Psi(\cdot)$ to the outcome. Examples of $\Psi(\cdot)$ include the hyperbolic tangent, the sigmoid, and rectified linear units (ReLU) activations. The MLP's output is an affine combination of the $K$ hidden outputs, where $\mathbf{v} =[v_j] \in \mathbb{R}^K $ is the vector of coefficients used in the linear combination and $b_1$ is the added bias term. 
	More precisely, the output of the MLP is given by:
	\begin{align}
		y_n=  \sum_{j=0}^{K-1}v_j\Psi\Bigg(\sum_{i = 0 }^{4M+1} w_{i,j}\mathbf{r}_{n, \text{vec}, i} + b_{0,j}\Bigg) + b_{1}. \label{MLP_Output_Equation}
	\end{align}
	The learnable parameters for the MLP are the weights $\mathcal{W}= \{\mathbf{W}, \mathbf{v}\}$ and biases $\mathcal{B}= \{\{b_{0,k}\}_{k=0}^{K-1},b_1\}$. Hence, the total number of parameters for the MLP is $|\mathcal{W}| + |\mathcal{B}|= 4MK+4K+1$. From \eqref{MLP_Output_Equation}, for each output sample, the MLP performs $4MK+3K$ multiplications and additions, and $K$ evaluations of $\Psi(\cdot)$.

	\subsection{Radial Basis Function Neural Network}
	
	\begin{figure}[t]
		\centering
		\includegraphics[width=0.4\textwidth]{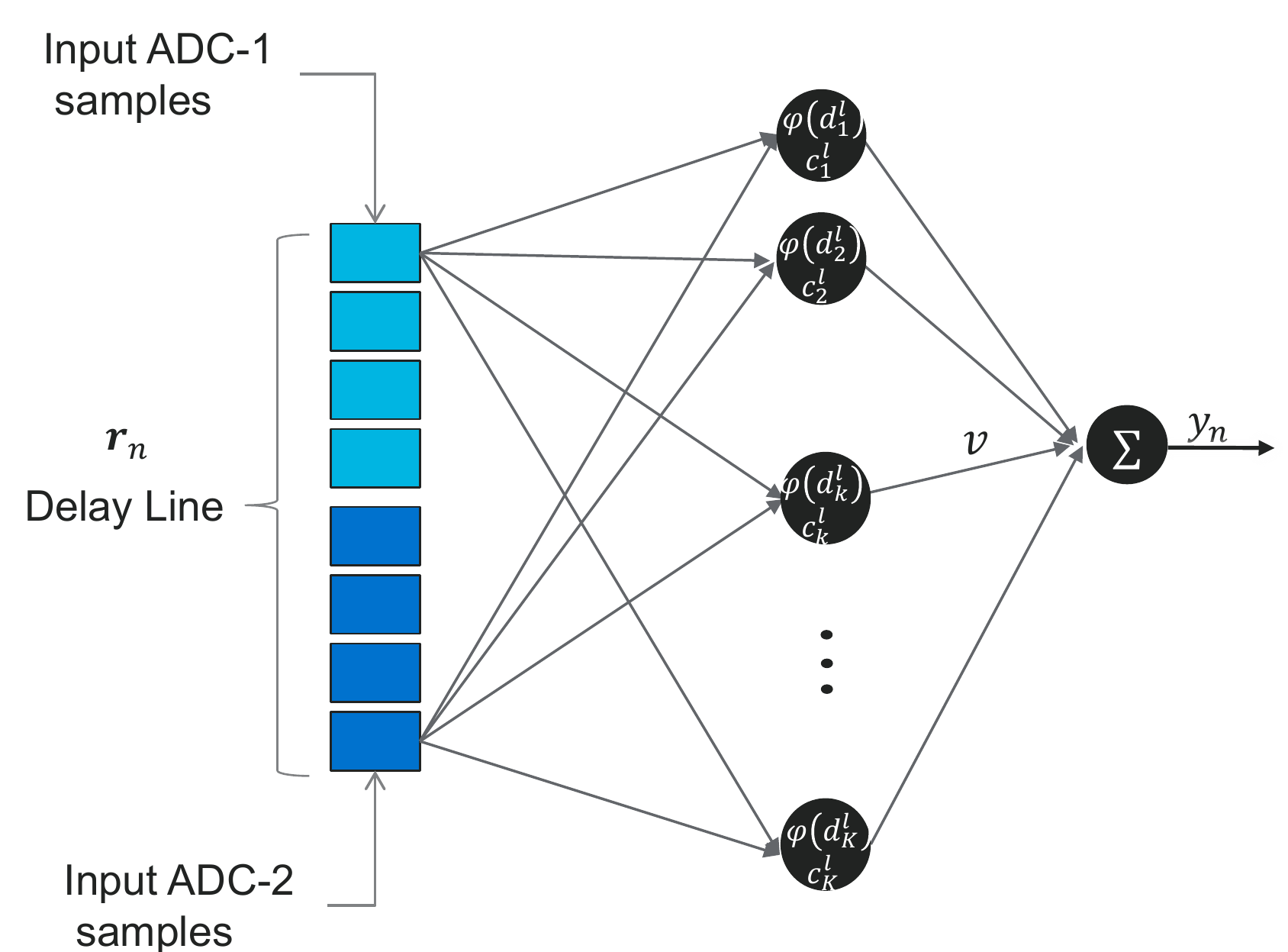}
		\caption{{\small Radial basis function neural network (RBFNN). The readback samples are used to compute the distances $d_k^l=\Vert \mathbf{r}_{n, \text{vec}} - \mathbf{c}_k^{l} \Vert$. A basis function $\phi(\cdot)$ is applied on the hidden outputs.}}
		\label{RBFNN_diagram}
	\end{figure}
	\begin{figure}[t]
		\centering
		\includegraphics[width=0.45\textwidth]{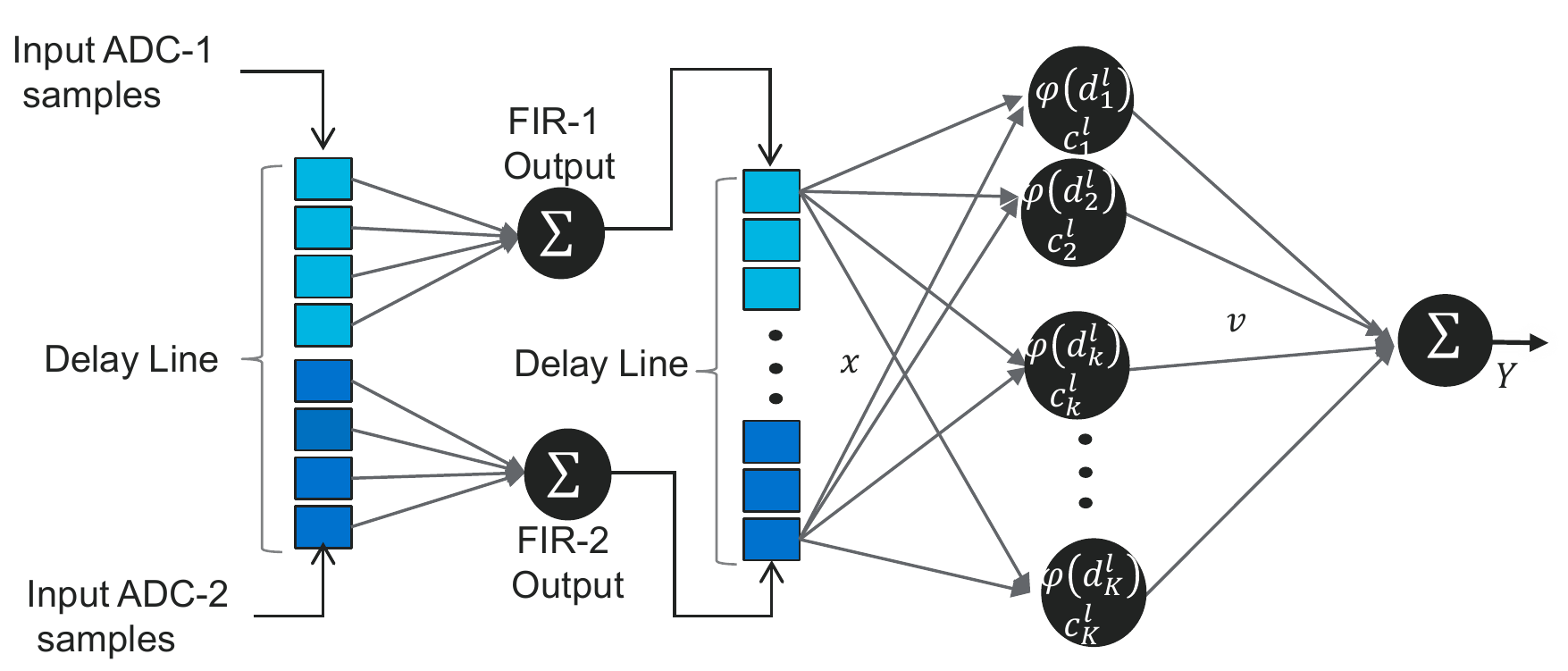}
		\caption{{\small Finite impulse response-radial basis function neural network (FIR-RBFNN). Two FIR filters reduce the input dimension so that the centroids $\mathbf{c}_k^l$ lie in a lower dimensional space compared with the readings $\mathbf{r}_n$, thereby reducing the overall complexity. }}
		\label{FIR_RBFNN_diagram}
	\end{figure}
	Fig.~\ref{RBFNN_diagram} shows the architecture of the RBFNN. The RBFNN replaces the matrix multiplication and non-linear activation in the MLP with distances from cluster centroids and a basis function, respectively. The RBFNN is characterized by the centroids $\mathbf{c}_k$ with $K$ biases $b_{0,k}$, $k=1,\ldots,K$, a length-$K$ vector $\mathbf{v}$ with bias $b_2$, a norm $\Vert\cdot\Vert$, typically the Euclidean norm, and a basis function $\phi(\cdot)$. The length of the centroids $\mathbf{c}_k$ matches the number of the input samples. The canonical implementation of the RBFNN uses the Euclidean distance. Given an input instance, the distance between the input samples vectors and centroids is computed. Then, the basis function is applied to the distances to provide the hidden output. Finally, the RBFNN output is computed as an affine combination of the hidden outputs. More specifically, the RBFNN output is described by:
	\begin{align}
		y_n= \sum_{k=0}^{K-1}v_k\phi(\Vert \mathbf{r}_{n, \text{vec}} - \mathbf{c}_k \Vert + b_{0,k} ) +b_1, \label{RBFNN_vect_equation}
	\end{align}
	where $\mathbf{c}_k\in \mathbb{R}^{4M+2}$. 
	To reduce the dimension of the centroids, we can use centroids that have the same length as the number of ADC samples per reader, instead of the vectorized ADC samples for both readers. If $K$ is even, then this formulation gives the output of the RBFNN as:
	\begin{align}
		y_n=  \sum_{l=0}^{1}\sum_{k=0}^{K/2-1}v_k\phi(\Vert \mathbf{r}_{n}^{\langle l\rangle} - \mathbf{c}_k^{l} \Vert + b_{l,k} )+b_l, \label{RBFNN_equation}
	\end{align}
	where $\mathbf{c}_k^{l}\in \mathbb{R}^{2M+1}$. In \eqref{RBFNN_equation}, $K/2$ centroids are assigned for each ADC sequence. The learnable parameters consist of the centers $\mathcal{C}= \{\mathbf{c}_k\}_{k=0}^{K-1}$, weights vector $\mathbf{v}$, and biases $\mathcal{B} = \{\{ b_{l,k} \}_{k=0}^{K-1}, b_l\}_{l=0}^1$. Thus, the RBFNN uses $|\mathcal{C}|+|\mathbf{v}|+|\mathcal{B}|=2KM+3K+2$ learnable parameters.
	
	
	\begin{figure}[t]
		\centering
		\includegraphics[width=0.42\textwidth]{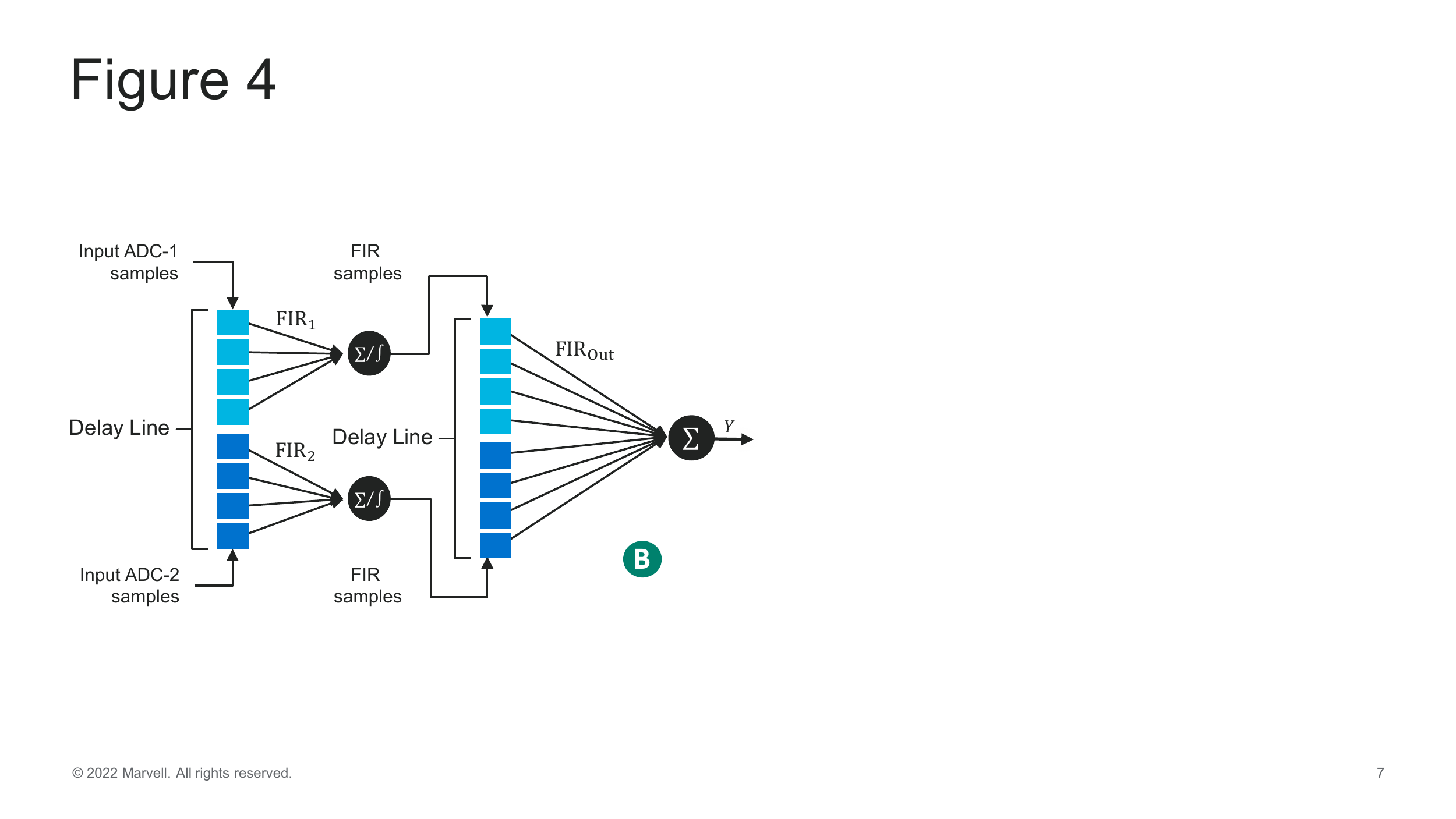}
		\caption{{\small RC-MLP1 architecture.}}
		\label{RC_MLP_Figure}
	\end{figure}
	\begin{figure}[t]
		\centering
		\includegraphics[width=0.42\textwidth]{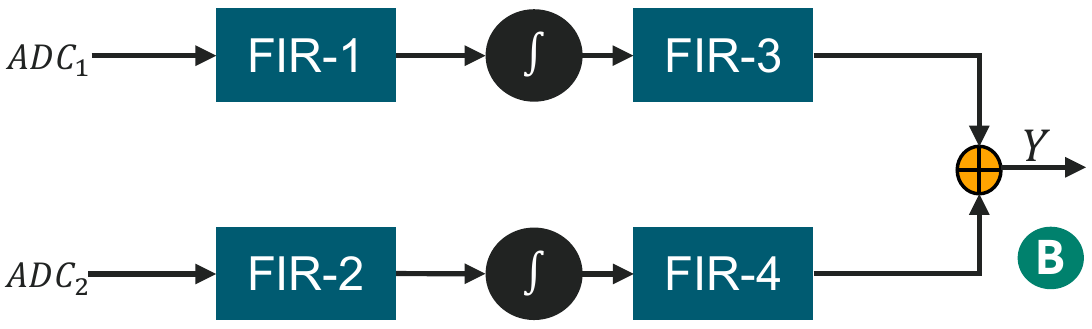}
		\caption{{\small FIR representation of RC-MLP1.}}
		\label{FIR_representation_RC_MLP}
	\end{figure}
	\begin{figure}[h]
		\centering
		\includegraphics[width=0.42\textwidth]{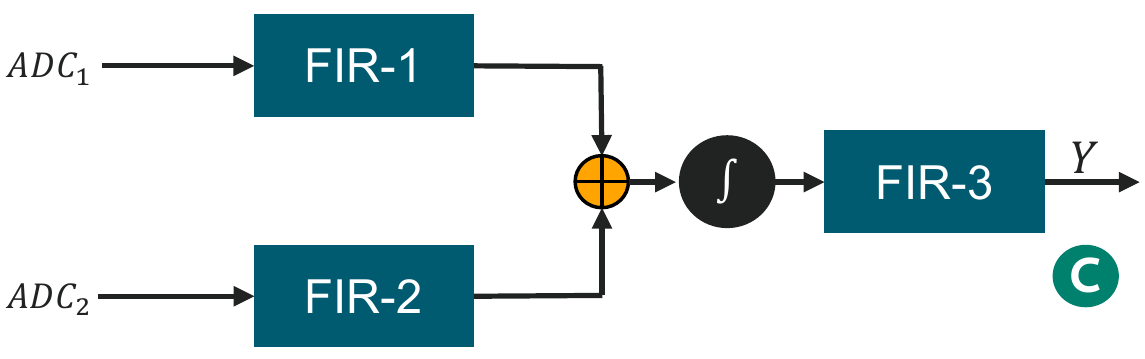}
		\caption{{\small RC-MLP2.}}
		\label{RC_MLP2_diagram}
	\end{figure}
	\begin{figure}[h]
		\centering
		\includegraphics[width=0.42\textwidth]{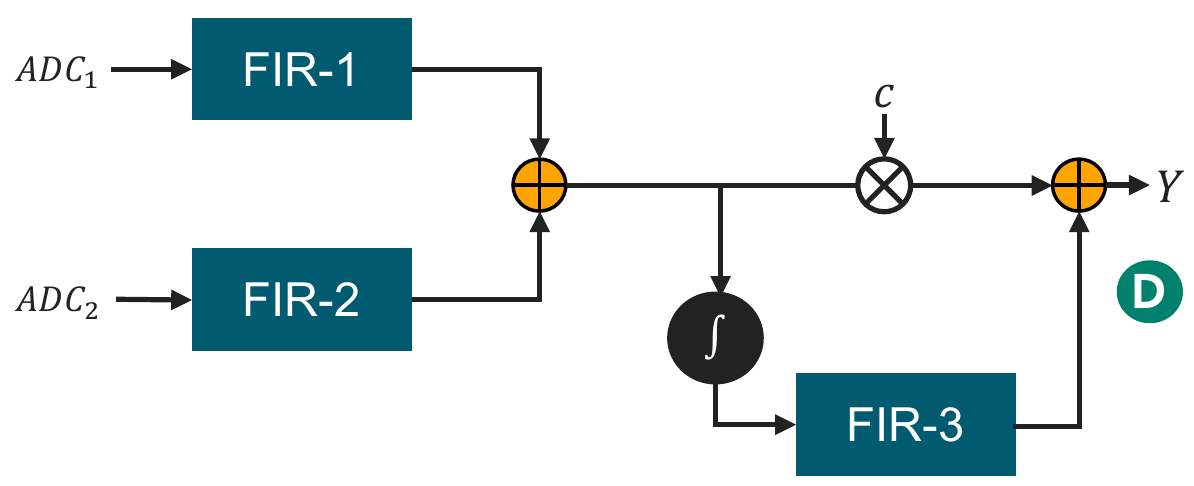}
		\caption{{\small RC-MLP3.}}
		\label{RC_MLP3_diagram}
	\end{figure}
	\begin{figure}[h]
		\centering
		\includegraphics[width=0.42\textwidth]{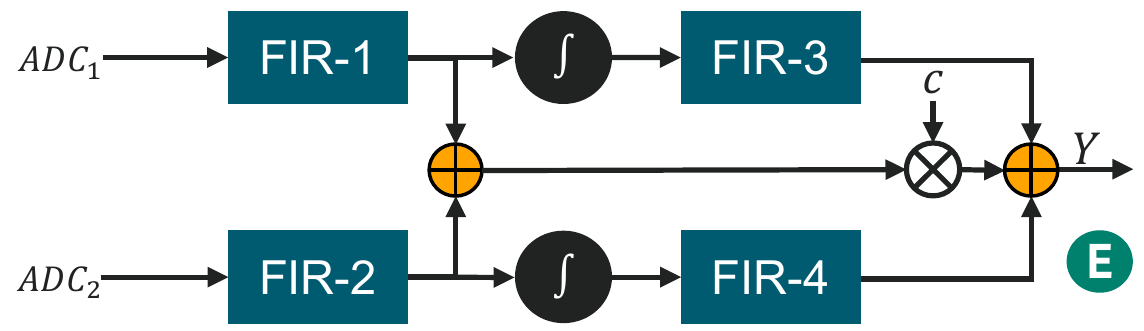}
		\caption{{\small RC-MLP4.}}
		\label{RC_MLP4_diagram}
	\end{figure}
	\section{Proposed Reduced Complexity Neural Network-based Equalizers}
	\label{ProposedNNEqualizers}
Many applications of NNs for computer vision utilize convolutional layers instead of fully connected matrix layers \cite{DeepLearning_Goodfellow}. In addition, convolutional neural networks have also been shown to be universal approximators \cite{zhou2020universalityofCNNs}. 

The authors of \cite{Data_Storage_Channel_Equalization_Neural_Networks_Nair, SimplifiedNonlinearEqualizers} observed that the rows of the trained matrix of weights in the MLP equalizer are approximately delayed replicas of one another for high density 1DMR. Therefore, the authors have proposed replacing the large matrix of weights in the MLP equalizer with an FIR filter.
	
	\subsection{FIR-RBFNN}
	To decrease the complexity of the RBFNN, we introduce an FIR filter $\mathbf{f}^{\langle l \rangle} \in \mathbf{R}^{2P+1}$, $l=0,1$, per ADC sequence, that interfaces the two input ADC sequences, as shown in Fig.~\ref{FIR_RBFNN_diagram}. The FIR filter maps the length-$(2M+1)$ ADC input to a length-$(2M'+1)$ sequence, where $M'<M$. The FIR-RBFNN output is given by:
	\begin{align}
		y_n= \sum_{l=0}^{1}\sum_{k=0}^{K/2-1}v_k\phi(\Vert (\mathbf{f}^{\langle l \rangle} * \mathbf{r}_{n}^{\langle l\rangle}) - \mathbf{c}_k^{l} \Vert + b_{l,k} )+b_l, \label{FIR_RBFNN_Equation}
	\end{align}
	where the dimension of the centroids is now $2M'+1$.
	\subsection{RC-MLP}
We propose four variants of reduced complexity MLP (RC-MLP) architectures. RC-MLP employs FIR filters to implement matrix multiplication and affine combination operations in MLP, which results in a considerably lower complexity architectures. The input layer introduces two FIR filters, FIR-1 and FIR-2, one for each stream of the ADC samples. According to the specific architecture, the hidden outputs corresponding to each filter are then stacked in one or two delay lines, as exhibited in Fig.~\ref{RC_MLP_Figure} and \ref{FIR_representation_RC_MLP}. An FIR filter can be viewed as a special case of a convolutional filter that performs only discrete-time convolutions.
	\subsubsection{RC-MLP1}
	Fig.~\ref{RC_MLP_Figure} shows the architecture of RC-MLP1. Two intermediary delay lines are introduced to store temporary hidden outcomes from FIR-1 and FIR-2 separately. The last layer consists of two FIR filters, FIR-3 and FIR-4, that map the delayed hidden samples to the final output.
	
	Let $\mathbf{f}^{\langle l \rangle}$ and $\mathbf{q}^{\langle l \rangle}$, $l=0,1$, denote the FIR filters interfacing the input ADC samples and hidden delayed outputs, respectively, shown in Fig. \ref{FIR_representation_RC_MLP}. The lengths of $\mathbf{f}^{\langle l \rangle}$ and $\mathbf{q}^{\langle l \rangle}$ are $2M+1$ and $K$, respectively, where $K$ is the number of hidden delay samples per ADC path. Then, the equalizer output $y_n$ is given in terms of the hidden outputs $h_n^{\langle l \rangle}$ as:
	\begin{align}
		&h_{n}^{\langle l \rangle}=\Psi((\mathbf{f}^{\langle l \rangle}*\mathbf{r}_n^{\langle l \rangle})_n + b_{0,l} )  \\
		&y_{n}=\sum_{l=0}^1 (\mathbf{q}^{\langle l\rangle} *\mathbf{h}_{n}^{\langle l \rangle})_n + b_{1},
	\end{align}
	where $b_{0,l}$ and $b_1$ are bias terms. RC-MLP1 uses $4M+2K+7$ learnable parameters and requires $4M + 2K + 2$ multipliers and two evaluations of $\psi(\cdot)$ per bit estimate.
	
	\subsubsection{RC-MLP2}
RC-MLP2 sums the outputs of FIR-1 and FIR-2. A nonlinear activation function is then applied to the sum. Consequently, only one hidden delay line with $K$ samples is required. FIR-3 combines the hidden delay line samples to provide the equalizer output. The hidden outputs and equalizer output can be expressed as:
	\begin{align}
		&h_{n}=\Psi\Bigg(\sum_{l=0}^1 ( \mathbf{f}^{\langle l \rangle}*\mathbf{r}_n^{\langle l \rangle})_n + b_{0} \Bigg)  \\
		&y_{n}=(\mathbf{q} *\mathbf{h}_{n})_n + b_{1}.
	\end{align}
	RC-MLP2 requires $4M+K+4$ parameters, $4M+K$ multipliers, and one evaluation of $\psi(\cdot)$ per bit estimate.
	
	\subsubsection{RC-MLP3}
The RC-MLP3 structure adds a linear connection between the linear output and the final output as can be seen in Fig.~\ref{RC_MLP3_diagram}. We added the linear connection in order to jump-start the system, using the FIR-1 and FIR-2 combination as a linear equalizer, while maintaining the improved generalization provided by the non-linear NN equalizer. As a result of the direct connection between the first hidden output and the final output, errors are more directly propagated from the output to the adapted FIR-1 and FIR-2. The equalizer's output for RC-MLP3 can be written as:
	\begin{align}
		&   h_{n,\text{ Linear}} = \sum_{l=0}^1 ( \mathbf{f}^{\langle l \rangle}*\mathbf{r}_n^{\langle l \rangle})_n + b_{0},\\
		&h_{n}= \Psi(h_{n,\text{ Linear}}) , \\
		&y_{n}= (\mathbf{q} *\mathbf{h}_{n})_n + ch_{n,\text{ Linear}}  + b_{1}.
	\end{align}
	
	\subsubsection{RC-MLP4}
	RC-MLP4 adds a linear connection to RC-MLP1 as shown in Fig.~\ref{RC_MLP4_diagram}. The output is given by:
	\begin{align}
		&   h_{n,\text{ Linear}}^{\langle l\rangle} =  ( \mathbf{f}^{\langle l \rangle}*\mathbf{r}_n^{\langle l \rangle})_n + b_{0,l}, \quad l=0,1,\\
		&h_{n}^{\langle l\rangle}= \Psi(h_{n,\text{ Linear}}^{\langle l\rangle}),  \\
		&y_{n}= \Big(\sum_{l=0}^1 \mathbf{q}^{\langle l\rangle} *\mathbf{h}_{n}^{\langle l\rangle}\Big)_n + c \sum_{l=0}^1 h_{n,\text{ Linear}}^{\langle l\rangle}  + b_{1}.
	\end{align}
	


	\section{Simulation Results}
	\begin{table}[t]
		\centering
		\caption{\small Performance and complexity comparison. The BER is computed over the first 20 sectors.} 
		\label{PerformancevsComplexity20Sectors}
		\setlength{\tabcolsep}{3pt}
			\begin{tabular}{@{}cccc@{}}
				\toprule
				Architecture & $K$ & BER  &Complexity\\ \midrule
				2D-LMMSE with fixed [3,7,1] target  &N/A    &0.027982 &22      \\
				\textbf{2D-LMMSE} &N/A            & \textbf{0.025548} &\textbf{22}\\
				2D-LECE           &N/A   & 0.023066 &22  \\
				2D-LECE with 21 Taps per ADC	        &N/A    &0.022658 &42      \\
				RBFNN              &6   &0.02315   &157     \\
				RBFNN             &20   &0.021608 &521       \\
				RBFNN           &30    &0.021733 &781        \\
				FIR-RBFNN, Gaussian Basis       &6    &0.021860 &107      \\
				FIR RBFNN, Tanh Basis    &6   &0.022497 &107        \\
				RC-FIR-RBFNN, Linear Basis    &6   &0.022773 &41        \\
				RC-FIR-RBFNN, Gaussian Basis   &6 &0.023744 &41    \\
				\textbf{MLP}             & 6   & \textbf{0.020757} &\textbf{145}      \\		
				RC-MLP1   &6       &0.022736 &31      \\ 
				RC-MLP1  & 10     &0.022048 &35 \\
				RC-MLP1 & 14      &0.021916 &39 \\
				RC-MLP1 & 18    &0.021862 &43 \\
				RC-MLP2 & 9      &0.021668 &34 \\
				\textbf{RC-MLP3}  & 9      &\textbf{0.021243} &\textbf{35} \\
				RC-MLP4  & 18    &0.021367 &44 \\
				\bottomrule
		\end{tabular}
	\end{table}

The performance and complexity of the discussed architectures have been assessed on raw ADC samples obtained from a HDD with TDMR technology. The data consists of $520$ sectors, where each sector contains about $40,000$ bits. The two read heads provide one ADC sample per bit, totaling two ADC samples per bit. The reader cross-track separation is $52\%$ of the $85$ nm track pitch.
	
	\subsection{Linear Equalizers - 20 Sectors}
A comparison of BER performances and complexity across the first 20 sectors is provided in Table 1. The equalizers' trainable parameters and target adapt to the CE criterion, except for the 2D-LMMSE. For linear equalizers, 11 taps per ADC sequence are used, unless otherwise noted. The computational complexity of a method is quantified by the number of learnable parameters it requires. The number of parameters is directly proportional to the number of elementary operations required per bit estimation. Target taps $g_i$ are optimized under a monic constraint in all experiments. A special case is an experiment with a fixed target of $[3,7,1]$ that uses 2D-LMMSE. In practice, such a configuration is used as a baseline. By adapting the target, the BER is reduced by $8.70 \%$ when the MSE is used as the criterion for adaptation. Furthermore, we consider a linear equalizer that is trained on cross-entropy (2D-LECE). Further reduction of the BER occurs when the linear equalizer is adapted with CE instead of MSE. By increasing the number of FIR taps (from $11$ taps to $21$ taps per ADC sequence), the linear equalizer's BER decreases by about $1.77\%$. Compared to the linear equalizer that has 11 taps per ADC, $21$ taps per ADC increases the linear equalizer's complexity by about $90\%$.
	
	\subsection{RBFNN-based Equalizers - 20 Sectors}
	The basic RBFNN equalizer increases complexity without significant improvement in the BER even with $20$ and $30$ cluster centroids per ADC.  
	To improve the performance, FIR-RBFNN is used. The two FIR filters, one per ADC sequence, interface the ADC samples and map the $11$ samples per ADC to $5$ samples per ADC. In this case, the centroids lie in $5$ dimensional space (instead of $11$), making the distance computation more efficient. This FIR-RBFNN architecture with $6$ centroids per ADC achieves BER performance that is only $1.67\%$ higher that the RBFNN with $20$ centroids, while requiring $4.87$ less parameters. Nonetheless, FIR-RBFNN's BER is $5.23\%$ lower than 2D-LECE, but its number of parameters is $4.86$ higher. 
	RC-FIR-RBFNN decreases complexity by about $37\%$ over FIR-RBFNN. With linear basis, RC-FIR-RBFNN achieves a similar BER as FIR-RBFNN. 
	\begin{figure}[t]
		\centering
		\includegraphics[width=0.49\textwidth]{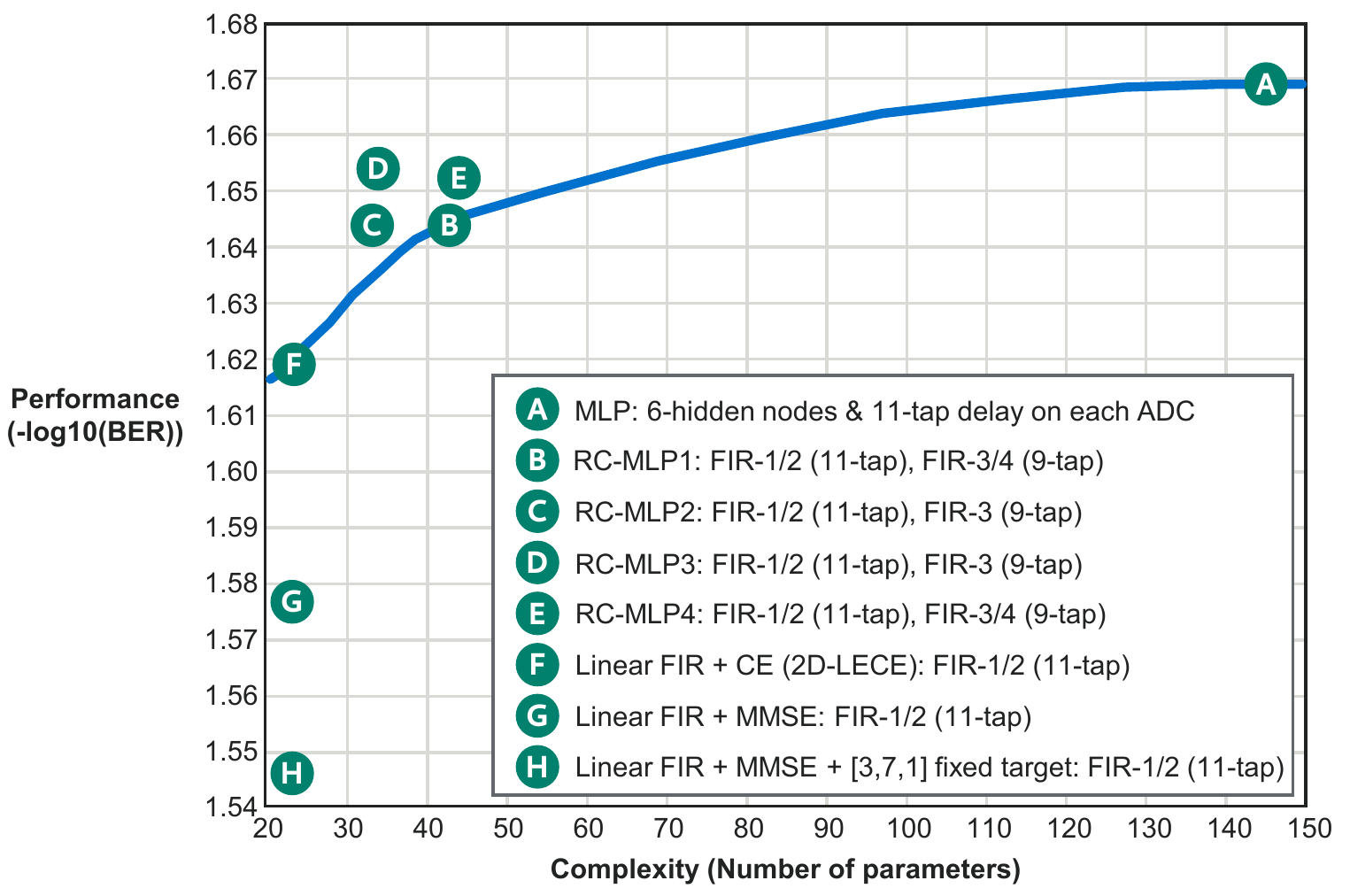}
		\caption{{\small Performance versus complexity for 520 sectors. Cross entropy is used for adaptation except when noted.}}
		\label{Performance_vs_complexity_520_sectors_plot}
	\end{figure}
	\subsection{RC-MLP - 20 Sectors}
We use the hyperbolic tangent function as the activation function $\Psi(\cdot)$ for MLP and RC-MLP. Positive and negative values of the noise-free PR signal $\hat{y}$ are equally likely. Hence, the activation function should span both the negative and positive real values symmetrically in MLP architectures with one hidden layer. MLP equalizers using the hyperbolic tangent activation achieved lower BERs than those using the rectified linear unit (ReLU) activation function as demonstrated in \cite{Nonlinear_Equalization_TDMR_Using_NNs_Shen}.

The MLP with $6$ hidden nodes achieves the lowest BER, which is $10\%$ lower than the 2D-LECE with $11$ taps while demanding a $6.6 \times$ increase in complexity. In comparison, RC-MLP1 with 6 hidden nodes achieves a BER performance that is $5.22\%$ lower than the 2D-LECE with only a $1.95 \times$ increase in complexity. RC-MLP2 achieves a $6.06\%$ lower BER than the 2D-LECE while requiring only a $1.54 \times$ increase in complexity. {\it RC-MLP3 achieves a BER reduction of $7.9\%$ with a complexity increase of $1.59 \times$ over the 2D-LECE.} RC-MLP4 achieves $7.37\%$ lower BER with a $2\times$ increase in complexity. Thus, RC-MLP3 achieves the best balance between complexity and performance improvement. Also, RC-MLP3 reduces the implementation complexity by about $4.14\times$ compared with the MLP.  MLP. Furthermore, RC-MLP3 reduces the BER by $24.08\%$ and $16.85\%$ compared to 2D-LMMSE with fixed and adapting targets, respectively. 

\subsection{Performance vs. Complexity on all 520 Sectors}	
Fig. \ref{Performance_vs_complexity_520_sectors_plot} summarizes the performance versus complexity for the linear, RC-MLP, and MLP equalizers when observed as an average over 520 sectors. In the case of MSE training of the linear equalizer, adapting the target results in an improvement by $6.68\%$ in BER. Adapting the linear equalizer coefficients and target coefficients based on CE rather than MSE further improves the performance by $9.26\%$. The RC-MLP and MLP equalizers are trained using CE. The MLP equalizer achieves a $10.91\%$ lower BER than the linear equalizer trained on CE, at the cost of increasing complexity by 6.6 times. In comparison, RC-MLP3 achieves a gain of $8.23\%$ over the linear equalizer trained on CE, at the cost of increasing complexity by 1.59 times.

\subsection{Mutual Information}
	\begin{table}[t]
	\centering
	\caption{\small Mutual information and BER computed on the first 40 sectors.} 
	\label{MutualInformationComparison}
	\setlength{\tabcolsep}{18pt}
	\begin{tabular}{@{}cccc@{}}
		\toprule
		Architecture & BER  &$I(\mathbf{\text{LLR}},\mathbf{u})$\\ \midrule
		2D-LMMSE		&0.025436	&0.89798 \\
		2D-LECE        &  0.023020 &0.90717  \\
		RC-MLP3	       &0.021196 &0.91436    \\
		\bottomrule
	\end{tabular}
\end{table}

It is instructive to examine whether the BER gain between the linear equalizer and RC-MLP3 translates into an increase in the mutual information between the LLRs and the written bits $\mathbf{u}$, which is denoted by $I(\mathbf{\text{LLR}},\mathbf{u})$. The mutual information is computed using the nominal algorithm in \cite{kraskov2004estimating, ross2014mutual}.

	
	Table \ref{MutualInformationComparison} summarizes the mutual information for both equalizers. We observe a significantly lower BER with RC-MLP3 over 2D-LECE. This improvement in BER also translates into an increase in the mutual information $I(\mathbf{\text{LLR}},\mathbf{u})$.
	For CE adaptation, adapting the parameters of the equalizers improves both the sign and magnitude of the LLRs because CE adaptation is a maximum likelihood adaptation \cite{Nonlinear_Equalization_TDMR_Using_NNs_Shen}. Consequently, we observe that both the BER and the mutual information have improved.
	
	\section{Conclusion}
We have examined the performance-complexity trade-off for different equalizer architectures. The multilayer perceptron (MLP) achieves significant performance gains over the linear equalizer. However, its complexity is about $6.6$ times that of the linear equalizer. We have proposed four variants of the reduced complexity MLP (RC-MLP) to facilitate practical implementation of non-linear neural network equalizer. The RC-MLP3 architecture outperforms its variants and achieves most of the performance gains of the MLP while requiring only $1.59\times$ the complexity of the linear equalizer. The complexities of MLP and RC-MLP scale as $O(LK+K)$ and $O(L+K)$, respectively, if $L$ is the length of the input and $K$ is the length of the hidden output. In future work, we will examine codeword error rates for different architectures.
	\bibliographystyle{IEEEtran}
	\bibliography{RCMLP_Bib}

\end{document}